\documentstyle[eqsecnum,aps,epsf]{revtex}            
\begin{document}
\draft
\preprint{Le/9609007}
\title{Quaternionic Electroweak Theory\\ and\\  
CKM matrix\footnote{Le/9609007,  
submitted for publication in Int.~J.~Theor.~Phys.}}
\author{Stefano De Leo\thanks{e-mail: {\em deleos@le.infn.it}} }
\address{Dipartimento di Fisica, Universit\`a degli Studi Lecce and INFN, 
Sezione di Lecce\\
via Arnesano, 73100 Lecce, Italy} 
\date{August, 1996}
\maketitle


\begin{abstract}

We find in our quaternionic version of the electroweak theory an apparently 
hopeless problem: In going from complex to quaternions, the calculation of 
the real-valued parameters of the CKM matrix drastically changes. We aim to 
explain this {\em quaternionic puzzle}.

\end{abstract}
\pacs{PACS number(s): 02.10.Tq, 02.20.Sv, 12.15.Ff; \\ 
      KeyWords: quaternions, CKC matrix, electroweak theory. }


\section{Introduction}
\label{s1}

In this paper we review some of the basic properties of the quaternionic 
electroweak theory~\cite{qet}, based on the one-dimensional local gauge 
group $U(1,q)_L \mid U(1,c)_Y$ (quaternionic counterpart of the Glashow 
group~\cite{gla}). Notwithstanding the recent success in manipulating the 
noncommutative quaternionic field in Quantum Mechanics and Field 
Theory~\cite{adl,adl1,adl2,del1,del2}, quaternions 
must be treated with prudence. For 
example we  meet with a puzzle in our version of the Salam-Weinberg 
model~\cite{sw}. Namely, {\em how to reproduce the right calculation of the 
real-valued parameters of the Cabibbo-Kobayashi-Maskawa (CKM) 
matrix}~\cite{ckm} {\em by quaternions.}

Historically, the quaternionic field was introduced by Hamilton~\cite{ham} in 
1843 and after the fundamental contributions to Quaternionic Quantum 
Mechanics by Finkelstein {\it et al.} ~\cite{fin1,fin2} (foundations 
of quaternionic quantum theories, quaternionic representations of 
compact groups, etc.) quaternions were somewhat an enigma for physicists. 
Quaternions was restored to life by the  work of 
Horwitz and Biedenharn~\cite{hb} (quaternionic tensor product, second 
quantization and gauge fields). In the Preface of the 
Adler's book~\cite{adl} we read:
{\it In particular, my decision to embark on a detailed investigation of 
quaternionic quantum mechanics arose both from a question posed to me by 
Frank Yang and from my study of a preliminary version of the 1984 paper by 
Larry Biedenharn and Larry Horwitz sent to me by the authors}. Today the 
Adler's book represents the main reference for the one who is working in 
such a research field.

Let us briefly discuss the features of the quaternionic numbers. The 
quaternionic algebra has been expounded in a series of papers~\cite{pap} and 
books~\cite{books}, the reader may refer to these for further details.  For 
convenience we repeat and develop the relevant points.

The quaternionic algebra over the real field $\cal R$ is a set
\begin{equation} 
\label{h} 
{\cal H} = \{\alpha + i\beta + j\gamma + k\delta ~\mid~ \alpha, \beta, 
\gamma, \delta \in {\cal R}\}
\end{equation}
with operation of multiplication defined according to the 
following rules for imaginary units:
\[i^{2} = j^{2} = k^{2} = -1\]
\[ij=k \; , jk=i \; , \;  ki=j \; ,\]
\[ji=-k \; , kj=-i \; , \;  ik=-j \; .\] 
In going from the complex numbers to the quaternions we lose the property of 
commutativity ($ij\neq ji)$. This represents a challenge in manipulating 
such a numeric field. 

Working with noncommutative numbers we must admit the existence of left and 
right multiplication, in fact the left action of the operator $\cal O$ on 
quaternions $q$
\[ {\cal O}q \]
is, in general, different from its right action
\[ q{\cal O}~. \]
In order to distinguish left/right actions we will use the following 
terminology for right acting operators
\[ (1\mid {\cal O}) \; q \equiv q {\cal O}~. \]
Namely, we introduce the concept of {\em barred} operators.

Among the favourable results in using {\em barred} operators we recall the 
possibility to reformulate Special Relativity by quaternions~\cite{rel}. 
Explicitly, the quaternionic generators of the Lorentz group are
\begin{eqnarray*}
\mbox{boost} \; \; (ct,x) \; \; & 
\; \; \frac{k \; \vert \; j - j \; \vert \; k}{2} \; \; ,\\
\mbox{boost} \; \; (ct,y) \; \; & 
\; \; \frac{i \; \vert \; k - k \; \vert \; i}{2} \; \; ,\\
\mbox{boost} \; \; (ct,z) \; \; & 
\; \; \frac{j \; \vert \; i - i \; \vert \; j}{2} \; \; ,\\
\mbox{rotation} \; \; around \; \; x \; \; & 
\; \; \; \;  \frac{i-1 \; \vert \; i}{2} \; \; ,\\
\mbox{rotation} \; \; around \; \; y \; \; & 
\; \; \; \;  \frac{j-1 \; \vert \; j}{2} \; \; ,\\
\mbox{rotation} \; \; around \; \; z \; \; & 
\; \; \; \;  \frac{k-1 \; \vert \; k}{2} \; \; .
\end{eqnarray*}
The four real quantities which identify the space-time point ($ct,x,y,z$) 
are represented by the quaternion
\[ q=ct + i x + j y + k z~.\]
This gives the natural generalization of the Hamilton's idea~\cite{ham}. 
The Irish physicist used quaternions to describe the rotations in the 
three-dimensional space
\begin{equation}
e^{(iu_x+ju_y+ku_z)\theta / 2} \; (ix+jy+kz) \; e^{-(iu_x+ju_y+ku_z)
\theta /2 }~,
\end{equation}
where, ${\bf u}\equiv (u_x,u_y,u_z)$ identifies the rotation axis, and 
$\theta$ the rotation angle.

This paper is structured as follows: In section~\ref{s2} we review some of the 
basic properties of the quaternionic electroweak theory, in particular we 
discuss complex scalar product, Dirac equation, complex projection of the 
Lagrangian, quaternionic Higgs field, one-dimensional local gauge group. In 
section~\ref{s3} we explain our {\em quaternionic puzzle}, by showing that 
in our quaternionic version of the Salam-Weinberg model, the CKM matrix must 
be ``complex-barred''. In the last section we draw our conclusions.

\section{Quaternionic Electroweak Theory}
\label{s2}

An essential ingredient in our version of Quaternionic Quantum Mechanics is 
what Rembieli\'nski~\cite{rem} called the adoption of a complex geometry 
(complex scalar products). This choice is certainly less ambitious than that of 
Adler~\cite{adl}, who advocates the use of a quaternionic geometry to 
reformulate a {\em new} quantum mechanics. Nevertheless we recall that up 
to a decade ago the use of quaternionic wave functions made the definition 
of tensor products ambiguous. Complex geometry allows us to overcome many 
problems due to the noncommutativity of quaternionic numbers.

Let us address the questions of whether and how Quaternionic Quantum 
Mechanics with complex geometry relates to the observed physical world. 

\subsection{Momentum operator in QQM}
\label{s2a}

Although there is in QQM an anti-self-adjoint operator, $\bbox{\partial}$, 
with all the properties of a translation operator, imposing a quaternionic 
geometry, there is no corresponding quaternionic self-adjoint operator with 
all the properties expected for a momentum operator. This hopeless situation 
is also highlighted in the Adler's book~\cite[pag.~63]{adl}. 

The usual choice ${\bf p}\equiv -i\bbox{\partial}$, still gives a 
self-adjoint operator with the standard commutation relations with the 
coordinates, but such an operator does not commute with the Hamiltonian, 
which will be in general, a quaternionic quantity. Nevertheless we can 
overcome such a difficulty using a complex scalar product 
\begin{equation}
\label{csp}
\langle \psi \mid  \varphi \rangle_c ~=~ \frac{1-i\mid i}{2}~
\langle \psi \mid  \varphi \rangle~,
\end{equation}
and defining as the appropriate momentum operator
\begin{equation}
\label{mo}
{\bf p} \equiv -\bbox{\partial} \mid i~.
\end{equation}
Now the momentum operator~(\ref{mo}) is {\em formally real} and so it 
commutes with a generic quaternionic Hamiltonian, more, by using a complex 
geometry, it represents a self-adjoint operators
\[
\langle \psi \mid  \bbox{\partial} \varphi i \rangle_c ~=~ 
\langle \bbox{\partial} \psi i \mid  \varphi \rangle_c~.
\]

Complex projections of scalar products were used by Horwitz and Biedenharn 
in order to obtain consistently multiparticle quaternionic 
states~\cite{hb}. In a recent paper~\cite{ten} we also find an explicit 
definition of quaternionic tensor product.

\subsection{Dirac equation and complex-valued Lagrangian}
\label{s2b}

An interesting application of quaternions in quantum physics is represented 
by the quaternionic formulation of the Dirac equation~\cite{dir}. The need 
to use complex scalar products no longer relies solely on arguments 
relative to tensor product spaces (multiparticle systems) but is explicit 
in the single free particle wave equation.

In order to write equations relativistically covariant , we must treat the 
space components and time in the same way, hence we are obliged to modify 
the standard equations by the following substitution
\[
i\partial_t ~ \rightarrow ~ \partial_t \mid i~,
\]
and so the first modification that must be made in rewriting the Dirac 
equation is
\begin{equation}
\label{de}
\partial_t \psi i = (\bbox{\alpha} \cdot {\bf p} + \beta m)\psi \quad \quad
[~{\bf p}\equiv -\bbox{\partial} \mid i~]~.
\end{equation} 
The Dirac algebra upon the {\em reals} (but not upon complex) has a two 
dimensional irreducible representation with quaternions. Thus the standard 
$4\times 4$ complex matrices $(\bbox{\alpha},\beta)$ reduce to $2\times 2$ 
quaternionic matrices. A particular representation is given by
\[
\beta=\left( \begin{array}{cc} 1 & 0\\ 0 & $-1$ \end{array} \right) \quad ,
\quad \bbox{\alpha} = {\bf Q} \left( \begin{array}{cc} 0 & 1\\ $-1$ & 0 
\end{array} \right) \quad  \quad [~{\bf Q}\equiv (i,j,k)~]~.
\]
Notwithstanding the two-component structure of the wave function, {\em all 
four standard solutions appear}
\[
u \; e^{-ipx} ~,~u \; je^{-ipx} ~,~v \; e^{-ipx} ~,~v \; je^{-ipx} ~. 
\] 
The trace theorems are modified~\cite{trace}, but the standard 
electrodynamics is reproduced. 

Let us now discuss the use of the variational principle within QQM. This is 
nontrivial because of the noncommutative nature of quaternions. As a first 
hypothesis we consider the traditional form for the Dirac-Lagrangian 
density:
\begin{equation}
\label{lag1}
{\cal L}=  \bar{\psi} \gamma^{\mu} \partial_{\mu} \psi i - m \bar{\psi} 
\psi~.
\end{equation}
The position of the imaginary unit is due to the fact that, in QQM, 
the $\partial_{\mu}$ operator is more 
precisely part of the first quantized momentum 
operator $\partial_{\mu}\mid i$. The previous Lagrangian is not hermitian, 
in fact
\[ (\bar{\psi} \gamma^{\mu} \partial_{\mu} \psi i)^{\dag} \neq 
\bar{\psi} \gamma^{\mu} \partial_{\mu} \psi i~.
\]
The correct form of the kinetic term reads:
\begin{equation}
\label{lag2}
{\cal L}_{k}=\frac{1}{2} \; [\bar{\psi}\gamma^{\mu}\partial_{\mu}\psi i-i
(\partial_{\mu}\bar{\psi})\gamma^{\mu}\psi] \; \; .
\end{equation}
This modification of eq.~(\ref{lag1}) yields hermitian our Lagrangian. 
The requirement of hermiticity 
however says nothing about the Dirac mass 
term in eq.~(\ref{lag1}). It is here that appeal to the variational 
principle must be made. A variation $\delta \psi$ in $\psi$ cannot in 
eq.~(\ref{lag2}) be brought to the extreme right because of the imaginary unit 
in the first half of the expression. The only consistent procedure is to 
generalize the variational rule that says that $\psi$ and $\bar{\psi}$ 
must be varied 
{\em independently}. We thus apply independent variations to $\psi$ 
 and $\psi i$, respectively  $\delta \psi$ and $\delta (\psi i)$. Similarly 
for $\delta \bar{\psi}$ and 
$\delta (i \bar{\psi})$. Now to obtain the desired Dirac equation for $\psi$ 
and its adjoint equation for $\bar{\psi}$ we are obliged to modify the mass 
term into
\begin{equation}
{\cal L}_{m}=-\frac{m}{2} \; [\bar{\psi} \psi-i\bar{\psi} \psi i] \; \; .
\end{equation}
The final result for $\cal L$ is~\cite{lag}
\begin{equation}
\label{lag3}
{\cal L}_{D}=\frac{1}{2} \; [\bar{\psi}\gamma^{\mu}\partial_{\mu}\psi i - i
(\partial_{\mu}\bar{\psi})\gamma^{\mu}\psi]-
\frac{m}{2} \; [\bar{\psi} \psi-i\bar{\psi} \psi i] \; \; .
\end{equation}
Considering this last equation we observe that it is nothing other than the 
complex projection of equation~(\ref{lag1})
\begin{equation}
\label{com}
{\cal L}_{D}=\frac{1-i\mid i}{2} \; {\cal L} \; \; .
\end{equation} 
Indeed, while $\cal L$ in eq.~(\ref{lag1}) is quaternionic and  
with the modification of ${\cal L}_{k}$ in eq.~(\ref{lag2}) hermitian, the 
form given in eq.~(\ref{lag3}) is purely complex and hermitian. Obviously 
we can write down a quaternionic hermitian Lagrangian and obtain the 
correct field equations through the {\em standard} variational principle by 
limiting $\delta \psi$ to complex variations (notwithstanding the 
quaternionic nature of the fields). We consider this latter options 
unjustified and thus select for the formal structure of $\cal L$ that of 
eq.~(\ref{lag3}).

\subsection{Doubling of solutions in bosonic equation}
\label{s2c}

The Dirac equation represents a {\em desirable} example of the so-called 
{\em doubling of solutions} in QQM with complex geometry. Obviously such a 
doubling of solutions occurs also in the bosonic equations. For example we 
find four complex orthogonal solutions for the Klein-Gordon equation, with 
the result that, in addition to the two normal solutions $e^{-ipx}$ (positive 
and negative energy), we discover two {\em anomalous} solutions 
$je^{-ipx}$. The 
physical significance of the anomalous solutions has been a ``puzzle'' for 
the authors. Only recently, by a quaternionic study of the electroweak 
Higgs sector~\cite{higgs}, we have been able to identify anomalous Higgs 
particles.

As remarked in the previous subsection, working within QQM, we need to 
generalize the variational principle. If $\varphi$ represents a quaternionic 
field its variations $\delta \varphi$ and 
$\delta (\varphi i )\equiv  i\delta \tilde{\varphi}$ must be treated 
independently. In fact we can vary $\varphi = \varphi_1 + j \varphi_2$ 
($\varphi_{1,2}$ complex) living $\tilde{\varphi} = \varphi_1 
- j \varphi_2$ unchanged. The complex projection also applies to 
the Klein-Gordon Lagrangian, explicitly
\[ {\cal L}_{KG} = (\partial_{\mu}  \varphi^{\dag} 
\partial^{\mu}  \varphi - m^2 \varphi^{\dag} \varphi)_c~,
\]
where 
\[ (\varphi^{\dag} \varphi )_c = \varphi_1^{\dag} \varphi_1  +
\varphi_2^{\dag} \varphi_2
\]
represents the quaternionic generalization of the standard term 
$\varphi_1^{\dag} \varphi_1$. The complex projection kills the 
{\em unpleasant} pure quaternionic cross term
\[ \varphi_1^{\dag} j \varphi_2 - \varphi_2^{\dag} j \varphi_1~.\]
To conclude this brief discussion on the quaternionic variational principle
(complete details are given in~\cite{lag,higgs}) we recall the main 
results found. {\em Going from complex to quaternionic fields we must admit 
quaternionic variations for our field, but only complex variations for our 
Lagrangian}. 

Since the only fundamental scalar could be the Higgs boson, in order to 
interpret the anomalous scalars we believe to be natural to concentrate our 
attention on the Higgs sector of the electroweak theory. Moreover the 
number of Higgs particles, before spontaneous symmetry breaking, is four 
and this agrees with the number of quaternionic solutions to the 
Klein-Gordon equation. The Higgs Lagrangian, in the quaternionic electroweak 
theory, is 
\begin{equation}
\label{higl}
 {\cal L}_{H} = (\partial_{\mu}  \varphi^{\dag} 
\partial^{\mu}  \varphi)_c  - \mu^2 (\varphi^{\dag} \varphi)_c - 
\vert \lambda \vert  (\varphi^{\dag} \varphi)_{c}^{2}~,
\end{equation}
where 
\[ \varphi \equiv h^{0} + j h^{+} \quad \quad [~h^0, \; h^{+} ~~~ 
\mbox{complex fields}~]~.
\]
The Lagrangian (\ref{higl}) is obviously invariant under the global group 
\[ U(1,q) \mid U(1,c)~,\]
quaternionic counterpart of the complex Glashow group
\[ SU(2,c) \times U(1,c)~.\]

\subsection{The quaternionic local gauge group}
\label{s2d}

We wish now to 
construct a fermionic Lagrangian invariant under the quaterninic group 
$U(1,q)$. If we consider a single particle (two component) field $\psi$, we 
have no hope to achieve this. In fact the most general transformation
\[ \psi \rightarrow f \psi g \quad \quad 
(f, \; g \; \mbox{quaternionic numbers} ) ~ , \]
is right-limited from the complex projection of our Lagrangian and 
left-limited 
from the presence of quaternionic (two dimensional) $\gamma^{\mu}$ matrices. 
So we could 
only write a Lagrangian invariant under a right-acting complex $U(1, \; c)$ 
group. The situation drastically changes if we use a ``left-real'' (four 
component) Dirac equation
\[(\tilde{\gamma}^{\mu}\partial_{\mu}\mid i-m) \psi = 0 ~,\]
where
\[ \tilde{\gamma}^{\mu} \equiv \gamma^{\mu} \mbox{-matrices with $i$-factors 
substituted by} \; 1\mid i ~ .
\]
In this case we could commute the quaternionic phase and restore 
the invariance under the left-acting quaternionic unitary group $U(1,q)$.

The massless fermionic Lagrangian (for the first generation) in our 
quaternionic electroweak model  reads
\begin{equation}
\label{fer}
{\cal L}_{c}^{f} = (\bar{\psi}_{l} \tilde{\gamma}^{\mu}\partial_{\mu} 
\psi_{l} i +
\bar{\psi}_{q} \tilde{\gamma}^{\mu}\partial_{\mu} \psi_{q} i )_{c} \; \; ,
\end{equation}
with
\[ \psi_{l}=e+j\nu \quad , \quad \psi_{q}=d+ju \quad \quad 
(e, \; \nu, \; d, \; u \; 
\mbox{complex fermionic fields})~ .\]
This Lagrangian is globally invariant under the following transformations:\\

\noindent 
\begin{tabular}{lrclc} 
-- left-handed fermions --~~~~~~~~~~~~~ &
$\quad e_{L}+j\nu_{L}$ & $\quad \rightarrow \quad$ & 
$e^{-\frac{g}{2} {\bf Q}\cdot \bbox{\alpha}} 
\; (e_{L}+j\nu_{L}) \; e^{\frac{\tilde{g}}{2} i Y_{l}^{(L)} \beta} \quad$ & ,\\
 & & & &  \\
 & $\quad d_{L}+ju_{L}$ & $\quad \rightarrow \quad$ & 
$e^{-\frac{g}{2} {\bf Q}\cdot \bbox{\alpha}} 
\; (d_{L}+ju_{L}) \; e^{\frac{\tilde{g}}{2} i Y_{q}^{(L)} \beta} \quad$ & ,\\
 & & & &  \\
-- right-handed fermions --~~~~~~~~~~~~~ &
$\quad e_{R}$ & $\quad \rightarrow \quad$ & $e_{R} \; 
e^{\frac{\tilde{g}}{2} i Y_{e}^{(R)} \beta}$ 
& , \\
 & & & &  \\
 & $\quad d_{R}+ju_{R}$  & $\quad \rightarrow \quad$ & $d_{R} \; 
e^{\frac{\tilde{g}}{2} i Y_{d}^{(R)} \beta} 
+ j u_{R} \; e^{\frac{\tilde{g}}{2} i Y_{u}^{(R)} \beta}$ & .\\
 & & & &  
\end{tabular}
\\

\noindent The {\em weak-hypercharge} assignments are 
\[ Y_{l}^{(L)}=-1 \quad , \quad Y_{q}^{(L)}=+\frac{1}{3} \quad , \quad 
Y_{e}^{(R)}=-2 \quad , \quad Y_{d}^{(R)}=-\frac{2}{3} \quad , \quad 
Y_{u}^{(R)}=+\frac{4}{3} \quad .\]
In order to construct a local-invariant theory, we must introduce the 
following quaternionic gauge field
\begin{equation}
W_{\mu}=W_{\mu}^{0}+jW_{\mu}^{+} 
\quad \quad [ \; W_{\mu}^{0}=(B_{\mu} +iW_{\mu}^{1})/\sqrt{2} 
\; \; , \; \; W_{\mu}^{+}=(W_{\mu}^{2}-iW_{\mu}^{3})/\sqrt{2} \; ]~ , 
\end{equation}

\begin{center}
\begin{tabular}{cclc}
${\bf W}^{\mu}$ & for & $U(1, \; q)_{L}$ & ,\\
$ B^{\mu}$ & for & $U(1, \; c)_{Y}$ & ,
\end{tabular}
\end{center}
by the covariant derivatives
\begin{center}
\begin{tabular}{lcl}
${\cal D}^{\mu} (e_{L}+j\nu_{L})$ & $\equiv$ & $[\partial^{\mu}-
\frac{g}{2} \; (i \mid W_{1}^{\mu}+ 
j \mid W_{2}^{\mu}+ 
k \mid W_{3}^{\mu})
- \frac{\tilde{g}}{2} \mid B^{\mu} i \; ](e_{L}+j\nu_{L}) ~,$\\
& & \\
${\cal D}^{\mu} (d_{L}+ju_{L})$ & $\equiv$ & $[\partial^{\mu} -
\frac{g}{2} \; (i \mid W_{1}^{\mu}+
j\mid W_{2}^{\mu}+
k\mid W_{3}^{\mu}) 
+ \frac{\tilde{g}}{6}  \mid B^{\mu} i \; ](d_{L}+ju_{L}) ~ ,$\\
& & \\
${\cal D}^{\mu} u_{R}$ & $\equiv$ & $(\partial^{\mu}
+ \frac{2\tilde{g}}{3} \mid B^{\mu}i) u_{R}  ~ ,$\\
& & \\
${\cal D}^{\mu} d_{R}$ & $\equiv$ & $(\partial^{\mu} 
- \frac{\tilde{g}}{3} \mid B^{\mu} i) d_{R}  ~ ,$\\
& & \\
${\cal D}^{\mu} e_{R}$ & $\equiv$ & $(\partial^{\mu} 
- {\tilde{g}} \mid B^{\mu} i)e_{R}  ~ ,$\\
 & &
\end{tabular}
\end{center}
\begin{center}
- the substitution $\partial^{\mu} \rightarrow {\cal D}^{\mu}$ in~(\ref{fer}) 
makes our Lagrangian locally invariant - .
\end{center}
Full details concerning gauge kinetic terms, Yukawa couplings, interactions 
among gauge bosons and fermions, symmetry breaking are reported 
elsewhere~\cite{qet}.

\section{The Quaternionic Puzzle}
\label{s3}

In our discussion of the electroweak model, we limited the number of 
fermion generations to one. We now lift that restriction and consider the 
implication of having $N$ generations. Although the existing experimental 
situation supports the value $N=3$, we shall take $N$ arbitrary in our 
analysis.

\subsection{Complex mixing matrix}
\label{s3a}

In this subsection we briefly recall the fermion mixing of 
the standard (complex) Salam-Weinberg model.  By convention, 
the mixing is assigned to the $Q=1/3$ quarks by
\begin{equation}
J_{ch}^{\mu} =  \bar{u}'_{L,\alpha} \gamma^{\mu} d'_{L,\alpha} 
             =  \bar{u}_{L,\alpha }  \gamma^{\mu} U^{\dag}_{L,\alpha \beta} 
                 D_{L,\beta \gamma} d_{L,\gamma} 
             =    \bar{u}_{L,\alpha} \gamma^{\mu} d''_{L,\alpha}~,
\end{equation}
where
\[ 
d''_{L,\alpha} = U^{\dag}_{L,\alpha \beta} D_{L,\beta \gamma} d_{L,\gamma}
              = V_{\alpha \gamma} d_{L,\gamma} \quad \quad 
(\alpha, \beta, \gamma = 1, \dots , N)~. 
\]
There is no difficulty in passing $U_L$ through $\gamma^\mu$ because the 
former matrix acts in flavor space whereas the latter matrices act in the 
spin space (obviously we also use the {\em commutation} of complex numbers). 

Thus the $Q=1/3$ quark states participating in transitions of the charged 
weak current are linear combinations of mass eigenstates. The quark-mixing 
matrix $V$, being the product of two unitary matrices, is itself unitary.

An $N\times N$ unitary (complex) matrix is characterized by $N^2$ 
real-valued parameters. Of these, $N(N-1)/2$ are angles and $N(N+1)/2$ are 
phases. Not all the phases have physical significance, because $2N-1$ of 
them can be removed by {\em quark rephasing}~\cite{dsm}. This leaves $V$ with
$(N-1)(N-2)/2$ such phases. Then, the unitary $N \times N$ (complex) matrix 
for $N$ quark generations possesses $(N-1)^2$ observable real parameters. 
Obviously, in going from complex to quaternions, the calculation of the 
real-valued parameters of the CKM matrix drastically changes.

\subsection{Complex-barred mixing matrix}
\label{s3b}

Working with quaternionic field (the primes signify that the states which 
appear in the original gauge-invariant Lagrangian are generally {\em not} 
the mass eigenstates) 
\[ \psi'_{L,\alpha}=d'_{L,\alpha}+j u'_{L,\alpha} \quad \quad 
(\alpha=1, \dots , N)~,\]
we must expect to have $N\times N$ barred-quaternionic unitary matrices 
instead of complex unitary matrices. 

Remembering that a barred-quaternion, in terms of real quantities, is 
expressed by
\[
{\cal Q} \equiv q + p \mid i \equiv \alpha_q + i \beta_q + j \gamma_q + k \delta_q + 
(\alpha_p + i \beta_p + j \gamma_p + k \delta_p ) \mid i~,
\]
with
\[ \alpha_{q,p}, \; \beta_{q,p}, \;  \gamma_{q,p}, \; \delta_{q,p} 
\in {\cal R}~,\]
we have
\[ \mbox{barred-quaternions} ~ \supset ~ \mbox{quaternions} ~ \supset ~
\mbox{complex} ~ , \]
and more
\[ \mbox{barred-quaternions} ~\supset ~ \mbox{barred-complex} 
~~[~\mbox{elements} \; \mbox{like} ~ \alpha + 
\beta  \mid i \equiv {\cal C}~] ~ .\]

We can now give the general formulas for counting the generators of unitary 
$N$-dimensional groups as a function of $N$:
\begin{center}
\begin{tabular}{lcccc} 
$U(N,  {\cal Q})$      & : & $4N+8 \; \frac{N(N-1)}{2}$ & = & $4N^{2}$~,  \\ 
$U(N,  q)$          & : & $3N+4 \; \frac{N(N-1)}{2}$ & = & $N(2N+1)$~, \\ 
$U(N,  {\cal C})$  & : &  $N+2 \; \frac{N(N-1)}{2}$ & = & $N^{2}$~. 
\end{tabular}
\end{center}
For a detailed discussion of quaternionic groups, the reader can consult 
the work of ref.~\cite{gut}.

Considering barred-quaternionic number as elements for our $N\times N$ 
unitary matrix we quadruple the real-valued parameters, consequently we 
increase the real-valued parameters counting of the CKM matrix. This 
{\em puzzle} is soon overcome by noting that the mixing matrix have to 
commute with the gauge group $U_{L}(1,q)\mid U_{Y}(1,c)$. This restriction 
reduces to barred-complex the elements of the matrices which mix 
left-handed quarks. Finally, we have
\[ \psi'_{L,\alpha}=   D_{L,\alpha \beta} d_{L,\beta}
                    +j U_{L,\alpha \beta} u_{L,\beta} \quad  \quad 
(\alpha, \; \beta =1, \dots , N)~,\]
with
\begin{center}
\begin{tabular}{ll}
$D_{L}, \; U_{L}$~~~~~~~~ &   unitary    barred-complex matrices,\\
$d_{L,R}, \; u_{L,R}$           & complex mass eigenstates.
\end{tabular}
\end{center}

The transformation from the gauge basis states to the mass basis states 
turns out to have no effect on the structure of the electromagnetic and 
neutral weak currents. As example of this, consider the quark contribution 
to the weak currents:
\begin{equation}
\left\{ [ \; \bar{d}_{L,\alpha} D^{\dag}_{L,\alpha \beta} -  
\bar{u}_{L,\alpha} 
U^{\dag}_{L,\alpha \beta} j \; ] \; \gamma_{\mu} \;  
[ \; -\frac{g}{2} \; ( \; i \mid W_{1}^{\mu}+
j\mid W_{2}^{\mu}+
k\mid W_{3}^{\mu} \; ) \; ] ~ [ \;  D_{L,\beta \gamma} d_{L,\gamma}
 +j U_{L,\beta \gamma} u_{L,\gamma} \; ] \right\}_c ~.
\end{equation}
After the complex projection we find no {\em flavor-changing neutral 
currents}, mixing between generations does 
manifest itself in the system of quark charged weak currents:
\[
-\frac{g}{2} \; \bar{d}_{L,\alpha} D^{\dag}_{L,\alpha \beta} \gamma_{\mu} 
\; ( \; j\mid W_{2}^{\mu}+
k\mid W_{3}^{\mu} \; ) \;  
j U_{L,\beta \gamma} u_{L,\gamma}  + ~~\mbox{h.c.}~, 
\]
and so
\begin{equation}
J_{ch}^{\mu} =  \bar{d}_{L,\alpha} \gamma_{\mu}
                 D^{\dag}_{L,\alpha \beta} 
                 U_{L,\beta \gamma} u_{L,\gamma} 
                 \; ( \;  W_{2}^{\mu} - i W_{3}^{\mu} \; ) + ~~ \mbox{h.c.}~
             =  \bar{d}_{L,\alpha} \gamma_{\mu}
                 V_{\alpha \gamma} 
                 u_{L,\gamma} 
                 \; ( \;  W_{2}^{\mu} - i W_{3}^{\mu} \; ) + ~~ \mbox{h.c.}~
\end{equation}
where
\[
V_{\alpha \gamma}= D^{\dag}_{L,\alpha \beta} 
                 U_{L,\beta \gamma}~. \]

\section{Conclusions}
\label{s4}

The primary interest of the author in recent years has been to demonstrate 
the possibility of using quaternions in the description of elementary 
particles. The complex projection of scalar products and Lagrangians 
represent the fundamental ingredient in reformulating Quaternionic Quantum 
Theories.

The noncommutative nature of quaternions made complicated the standard 
approach to physical world. A complex geometry seems necessary (if not 
sufficient) for reproducing standard Quantum Theories. In this work we have 
reviewed the Quaternionic Electroweak Theory, based on the one-dimensional 
local gauge group $U_{L}(1,q) \mid U_{Y}(1,c)$ (minimal quaternionic 
unitary group for our Lagrangian) and overcome the apparent {\em puzzle} 
concerning the calculation of the real-valued parameters of the CKM matrix.

\end{document}